\documentclass{revtex4}
\usepackage{amsthm,amssymb,amsmath}				
\usepackage{graphicx,comment}
\usepackage{float}   
\usepackage{xcolor}

\begin{document}

\title{On the Klein-Gordon G\"{u}rses-oscillators and pseudo-G\"{u}rses-oscillators: vorticity-energy
correlations and spacetime associated degeneracies}
\author{Omar Mustafa}
\email{omar.mustafa@emu.edu.tr}
\affiliation{Department of Physics, Eastern Mediterranean University, G. Magusa, north
Cyprus, Mersin 10 - Turkey.}

\begin{abstract}
\textbf{Abstract:}\ We discuss KG-oscillators in the (1+2)-dimensional G\"{u}rses spacetime and under position-dependent mass (PDM) settings. We observe that the KG-G\"{u}rses oscillators are introduced as a byproduct of the very nature of the G\"{u}rses spacetime structure. We report that the energy levels of such KG-G\"{u}rses oscillators admit vorticity-energy correlations as well as spacetime associated degeneracies (STAD). We discuss KG-G\"{u}rses oscillators' results reported by Ahmed \cite{Ahmed1 2019} and pinpoint
his improper treatment of this model so that his results should be redirected to those reported in this study. Moreover, we introduce a new set of KG pseudo-G\"{u}rses oscillators that admits isospectrality and invariance with the KG-G\"{u}rses oscillators and inherits the same vorticity-energy correlations as well as STADs.


\end{abstract}

\maketitle

\section{Introduction}

Klein-Gordon (KG) and Dirac oscillators \cite{Moshinsky 1989,Bruce
1993,Dvoeg 1994,Mirza 2004} have received much attention over the years.
KG-oscillators in G\"{o}del-type spacetime (e.g., \cite{Moshinsky 1989,Bruce
1993,Dvoeg 1994,Das 2008,Carvalho 2016,Garcia 2017,Vitoria 2016}), in cosmic
string spacetime and Kaluza-Klein theory backgrounds (e.g., \cite{Ahmed1
2021,Boumal 2014}), in Minkowski spacetime with space-like dislocation \cite%
{Mustafa1 2022}, in Som-Raychaudhuri \cite{Wang 2015}, in (1+2)-dimensional G%
\"{u}rses space-time backgrounds (e.g., \cite{Gurses 1994,Ahmed1
2019,Mustafa2 2022}). The KG-oscillators in a (1+2)-dimensional G\"{u}rses
spacetime described by the metric%
\begin{equation}
ds^{2}=-dt^{2}+dr^{2}-2\Omega r^{2}dtd\theta +r^{2}\left( 1-\Omega
^{2}r^{2}\right) d\theta ^{2}=g_{\mu \nu }dx^{\mu }dx^{\nu };\text{ }\mu
,\nu =0,1,2,  \label{e1}
\end{equation}%
were investigated investigated by Ahmed \cite{Ahmed1 2019}., using $%
a_{_{0}}=b_{_{0}}=e_{_{0}}=1$, $b_{_{1}}=c_{_{0}}=\lambda _{_{0}}=0$, and
vorticity $\Omega =-\mu /3$, in the G\"{u}rses metric%
\begin{equation}
ds^{2}=-\phi dt^{2}+2qdtd\theta +\frac{h^{2}\psi -q^{2}}{a_{_{0}}}d\theta
^{2}+\frac{1}{\psi }dr^{2}  \label{e2}
\end{equation}%
(i.e., as in Eq.(5) of \cite{Gurses 1994}) where%
\begin{equation}
\phi =a_{_{0}},\,\psi =b_{_{0}}+\frac{b_{_{1}}}{r^{2}}+\frac{3\lambda _{_{0}}%
}{4}r^{2},\,q=c_{_{0}}+\frac{e_{_{0}}\mu }{3}r^{2},\,h=e_{_{0}}r,\,\lambda
_{_{0}}=\lambda +\frac{\mu ^{2}}{27}.  \label{e3}
\end{equation}%
In this note, we shall show that there are more quantum mechanical
features indulged in the spectroscopic structure of the KG-oscillators in
the background of such a G\"{u}rses spacetime metric (\ref{e1}) than those
reported by Ahmed \cite{Ahmed1 2019}, should this model be properly
addressed. Throughout this note, such KG-oscillators shall be called KG-G\"{u}rses oscillators.

We organize the current note in the following manner. In section 2, we
revisit KG-oscillators in the (1+2)-dimensional G\"{u}rses spacetime of (\ref{e1}) and present them in a more general form, that includes
position-dependent mass (PDM, which is a metaphoric notion) settings along
with Mirza-Mohadesi's KG-oscillators \cite{Mirza 2004} recipe. We observe
that the KG-G\"{u}rses oscillators are introduced as a byproduct of the very
nature of the G\"{u}rses spacetime structure. This motivates us to first
elaborate and discuss, in section 3, the effects of G\"{u}rses spacetime on
the energy levels of the KG-G\"{u}rses oscillators, without the
KG-oscillator prescription of Mirza-Mohadesi \cite{Mirza 2004}. Therein, we report that such KG-G\"{u}rses oscillators admit
vorticity-energy correlations as well as spacetime associated degeneracies
(STADs). In section 4, we discuss Ahmed's model \cite{Ahmed1 2019} that
includes Mirza-Mohadesi \cite{Mirza 2004} recipe and pinpoint Ahmed's \cite{Ahmed1 2019} improper treatment of the model at
hand. We consider the PDM KG-G\"{u}rses oscillators in section 5. We discuss
and report KG pseudo-G\"{u}rses oscillators in section 6, where we observe that they admit isospectrality and invariance with the KG G\"{u}rses-oscillators and inherit the same vorticity-energy correlations as well as STADs. Our concluding
remarks are given in section 7.

\section{KG-G\"{u}rses oscillators and PDM settings}

The covariant and contravariant metric tensors corresponding to the
(1+2)-dimensional G\"{u}rses spacetime of (\ref{e1}), respectively, read%
\begin{equation}
g_{\mu \nu }=\left( 
\begin{tabular}{ccc}
$-1\smallskip $ & $0$ & $-\Omega r^{2}$ \\ 
$0$ & $1\smallskip $ & $0$ \\ 
$-\Omega r^{2}$ & $\,0$ & $\,r^{2}\left( 1-\Omega ^{2}r^{2}\right) $%
\end{tabular}%
\right) \Longleftrightarrow g^{\mu \nu }=\left( 
\begin{tabular}{ccc}
$\left( \Omega ^{2}r^{2}-1\right) $ & $0\smallskip $ & $-\Omega $ \\ 
$0$ & $1\smallskip $ & $0$ \\ 
$-\Omega $ & $\,0$ & $\,1/r^{2}$%
\end{tabular}%
\right) \text{ };\text{ \ }\det \left( g_{\mu \nu }\right) =-r^{2}.
\label{e4}
\end{equation}%
Then the corresponding KG-equation is given by%
\begin{equation}
\frac{1}{\sqrt{-g}}\partial _{\mu }\left( \sqrt{-g}g^{\mu \nu }\partial
_{\nu }\Psi \right) =m^{2}\Psi .  \label{e5}
\end{equation}%
However, we shall now use the momentum operator%
\begin{equation}
p_{\mu }\longrightarrow p_{\mu }+i\mathcal{F}_{\mu },  \label{e6}
\end{equation}%
so that it incorporates the KG-oscillator prescription of Mirza-Mohadesi 
\cite{Mirza 2004} as well as position-dependent mass (PDM) settings proposed
by Mustafa \cite{Mustafa1 2022}. Where $\mathcal{F}_{\mu }=\left( 0,\mathcal{%
F}_{r},0\right) $ and our $\mathcal{F}_{r}=\eta r;$ $\eta =m\omega ,$ of 
\cite{Ahmed1 2019} and $\mathcal{F}_{r}=\eta r+g^{\prime }\left( r\right)
/4g\left( r\right) $ to also include PDM settings as in Mustafa \cite{Mustafa1 2022}. This would suggest that Ahmed's model is retrieved when the
positive-valued scalar multiplier $g\left( r\right) =1$. Nevertheless, the
reader should be aware that the regular momentum operator $p_{\mu }$ is
replaced by the PDM-momentum operator $p_{\mu }+i\mathcal{F}_{\mu }$ to
describe PDM KG-particles in general (for more details on this issue the
reader is advised to refer to \cite{Mustafa1 2022}).

Under such assumptions, the KG-equation (\ref{e5}) would transform into%
\begin{equation}
\frac{1}{\sqrt{-g}}\left( \partial _{\mu }+\mathcal{F}_{\mu }\right) \left[ 
\sqrt{-g}g^{\mu \nu }\left( \partial _{\nu }-\mathcal{F}_{\nu }\right) \Psi %
\right] =m^{2}\Psi ,  \label{e7}
\end{equation}%
which consequently yields%
\begin{equation}
\left\{ -\partial _{t}^{2}+\left( \Omega \,r\,\partial _{t}-\frac{1}{r}%
\partial _{\theta }\right) ^{2}+\partial _{r}^{2}+\frac{1}{r}\partial
_{r}-M\left( r\right) -m^{2}\right\} \Psi =0,  \label{e8}
\end{equation}%
where%
\begin{equation}
M\left( r\right) =\frac{\mathcal{F}_{r}}{r}+\mathcal{F}_{r}^{\prime }+%
\mathcal{F}_{r}^{2}.  \label{e9}
\end{equation}%
We now substitute%
\begin{equation}
\Psi \left( t,r,\theta \right) =\exp \left( i\left[ \ell \theta -Et\right]
\right) \psi \left( r\right) =\exp \left( -i\left[ \ell \theta -Et\right]
\right) \frac{R\left( r\right) }{\sqrt{r}}  \label{e10}
\end{equation}%
to imply 
\begin{equation}
R^{\prime \prime }\left( r\right) +\left[ \lambda -\frac{\left( \ell
^{2}-1/4\right) }{r^{2}}-\tilde{\omega}^{2}r^{2}-\tilde{M}\left( r\right) %
\right] R\left( r\right) =0,  \label{e11}
\end{equation}%
where $\ell =0,\pm 1,\pm 2,\cdots $ is the magnetic quantum number,%
\begin{equation}
\tilde{M}\left( r\right) =-\frac{3}{16}\left( \frac{g^{\prime }\left(
r\right) }{g\left( r\right) }\right) ^{2}+\frac{1}{4}\frac{g^{^{\prime
\prime }}\left( r\right) }{g\left( r\right) }+\frac{1}{4}\frac{g^{\prime
}\left( r\right) }{rg\left( r\right) }+\frac{1}{2}\frac{g^{\prime }\left(
r\right) }{g\left( r\right) }\eta r,  \label{e12}
\end{equation}%
and%
\begin{equation}
\lambda =E^{2}-2\,\Omega \,\ell \,E-2\eta -m^{2}\text{ ; \ }\tilde{\omega}%
^{2}=\Omega ^{2}E^{2}+\eta ^{2}.  \label{e13}
\end{equation}%
It is obvious that we retrieve Ahmed's model \cite{Ahmed1 2019} when $%
g\left( r\right) =1$. Moreover, we observe that the KG-G\"{u}rses oscillators are introduced as a byproduct of the very nature of the G\"{u}rses spacetime structure. This motivates us to first elaborate and discuss the effects of G\"{u}rses spacetime on the energy levels of the KG-G\"{u}rses oscillators, without the KG-oscillator prescription of Mirza-Mohadesi \cite{Mirza 2004} (i.e., with $\eta =0$). 
\section{KG-G\"{u}rses oscillators: vorticity-energy correlations and spacetime associated degeneracies}

It is obvious that KG-G\"{u}rses oscillators are introduced by the very
structure of G\"{u}rses spacetime. That is, for $\eta =0$, and $g\left(
r\right) =1$ our KG-equation (\ref{e11}) collapses into the two-dimensional
Schr\"{o}dinger oscillator%
\begin{equation}
R^{\prime \prime }\left( r\right) +\left[ \lambda -\frac{\left( \ell
^{2}-1/4\right) }{r^{2}}-\Omega ^{2}E^{2}r^{2}\right] R\left( r\right) =0,
\label{e14}
\end{equation}%
which admits exact textbook solvability so that the eigenvalues and radial
eigenfunctions, respectively, read%
\begin{equation}
\lambda =2\left\vert \Omega E\right\vert \left( 2n_{r}+\left\vert \ell
\right\vert +1\right)   \label{e15}
\end{equation}%
and 
\begin{equation}
R\left( r\right) \sim r^{\left\vert \ell \right\vert +1/2}\exp \left( -\frac{%
\left\vert \Omega E\right\vert r^{2}}{2}\right) L_{n_{n}}^{\left\vert \ell
\right\vert }\left( \left\vert \Omega E\right\vert r^{2}\right)
\Longleftrightarrow \psi \left( r\right) \sim r^{\left\vert \ell \right\vert
}\exp \left( -\frac{\left\vert \Omega E\right\vert r^{2}}{2}\right)
L_{n_{n}}^{\left\vert \ell \right\vert }\left( \left\vert \Omega
E\right\vert r^{2}\right) .  \label{e16}
\end{equation}%
where $L_{n_{n}}^{\left\vert \ell \right\vert }\left( \left\vert \Omega
E\right\vert r^{2}\right) $ are the generalized Laguerre polynomials. Now
with the help of (\ref{e13}) and (\ref{e15}) we obtain%
\begin{equation}
E^{2}-2\,\Omega \,\ell \,E-m^{2}=2\left\vert \Omega E\right\vert \left(
2n_{r}+\left\vert \ell \right\vert +1\right) .  \label{e16.1}
\end{equation}%
\begin{figure}[!ht!]  
\centering
\includegraphics[width=0.3\textwidth]{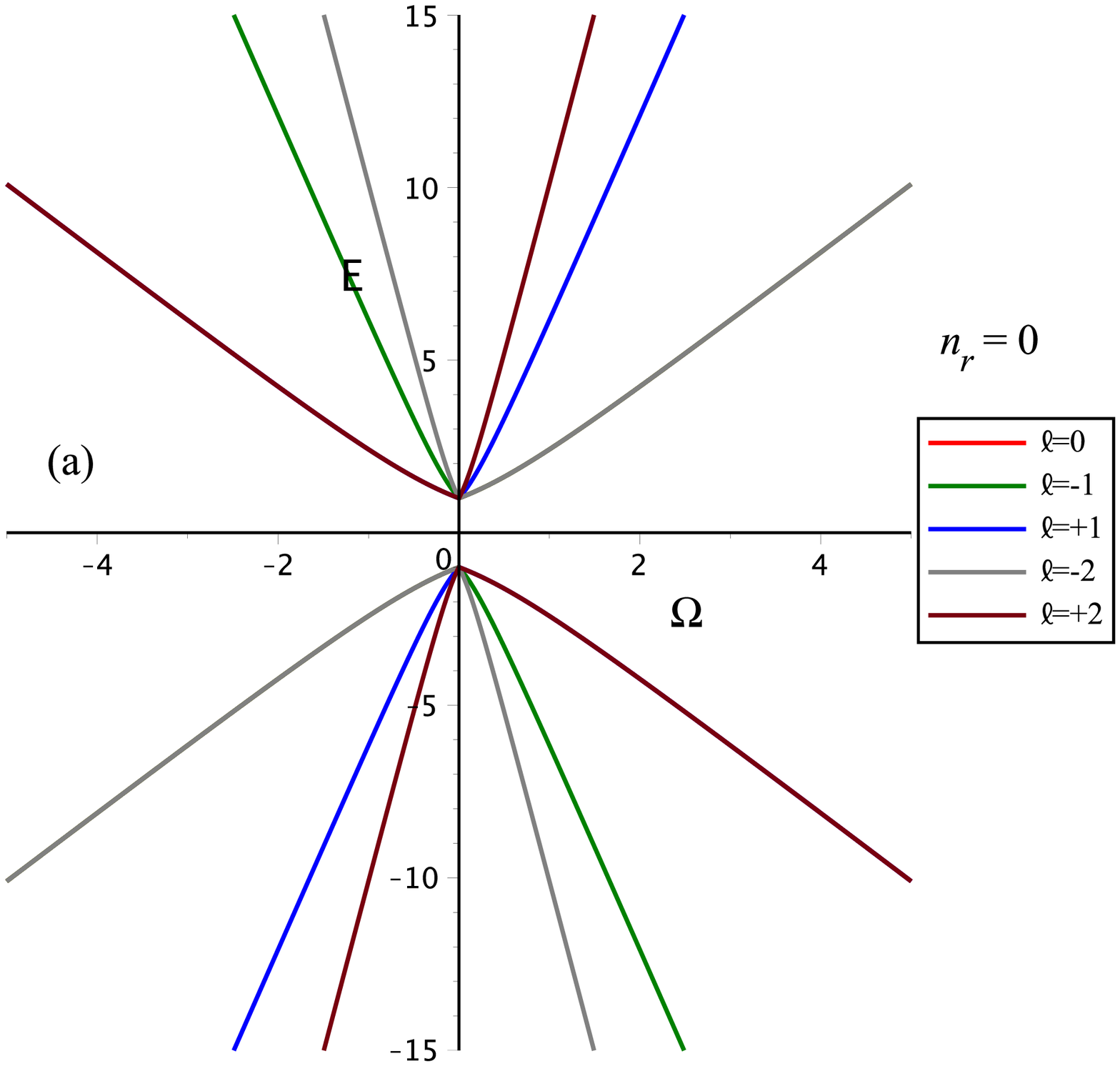}
\includegraphics[width=0.3\textwidth]{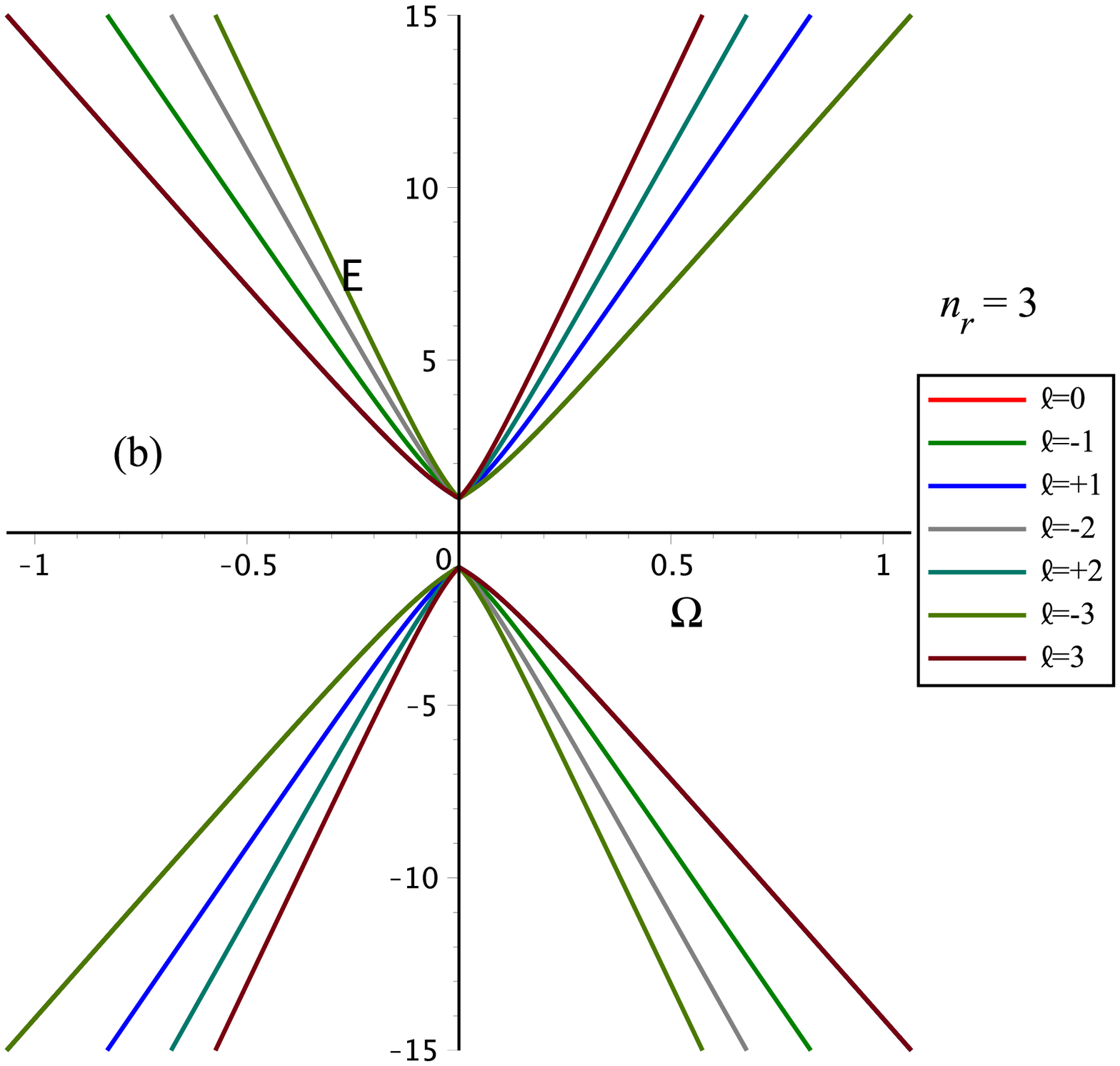}
\caption{\small 
{ The energy levels for KG-G\"{u}rses oscillators of (\ref{e16.4}) and (\ref{e16.5}) are plotted  $m=1$ (a) for $n_r=0, \, \ell=0,\pm1,\pm2$, and (b) for  $n_{r}=3$, $\ell=0,\pm1, \pm2,\pm3$.}}
\label{fig1}
\end{figure}%
This result should be dealt with diligently and rigorously, as mandated by the very nature of $\left\vert \Omega E\right\vert =\Omega _{\pm }E_{\pm }\geq 0$ or $%
\left\vert \Omega E\right\vert =-\Omega _{\mp }E_{\pm }\geq 0$ (that secures the finiteness and square integrability of the radial wavefunction (\ref{e16})), where $\Omega _{\pm }=\pm \left\vert \Omega \right\vert $ and $E_{\pm
}=\pm \left\vert E\right\vert $, That is, for $\left\vert \Omega
E\right\vert =\Omega _{\pm }E_{\pm }$ in (\ref{e16.1}) we obtain%
\begin{equation}
E_{\pm }^{2}-2\,\Omega _{\pm }E_{\pm }\,\tilde{n}_{+}-m^{2}=0;\;\tilde{n}%
_{+}=2n_{r}+\left\vert \ell \right\vert +\ell \,+1,  \label{e16.2}
\end{equation}%
and for $\left\vert \Omega E\right\vert =-\Omega _{\mp }E_{\pm }$ we get%
\begin{equation}
E_{\pm }^{2}+2\,\Omega _{\mp }E_{\pm }\,\tilde{n}_{-}\,-m^{2}=0;\;\tilde{n}%
_{-}=2n_{r}+\left\vert \ell \right\vert -\ell \,+1.  \label{e16.3}
\end{equation}%
Which would allow us to cast%
\begin{equation}
E_{\pm }=\Omega _{\pm }\,\tilde{n}_{+}\pm \sqrt{\Omega ^{2}\tilde{n}%
_{+}^{2}+m^{2}}\Rightarrow \left\{ 
\begin{tabular}{l}
$E_{+}=\Omega _{\pm }\,\tilde{n}_{+}+\sqrt{\Omega ^{2}\tilde{n}_{+}^{2}+m^{2}%
}$ \\ 
$E_{-}=\Omega _{\pm }\,\tilde{n}_{+}-\sqrt{\Omega ^{2}\tilde{n}_{+}^{2}+m^{2}%
}$%
\end{tabular}%
\right. ,  \label{e16.4}
\end{equation}%
for $\left\vert \Omega E\right\vert =\Omega _{\pm }E_{\pm }$ and%
\begin{equation}
E_{\pm }=-\Omega _{\mp \,}\tilde{n}_{-}\,\pm \sqrt{\Omega ^{2}\tilde{n}%
_{-}^{2}+m^{2}}\Rightarrow \left\{ 
\begin{tabular}{l}
$E_{+}=-\Omega _{-\,}\tilde{n}_{-}\,+\sqrt{\Omega ^{2}\tilde{n}_{-}^{2}+m^{2}%
}$ \\ 
$E_{-}=-\Omega _{+\,}\tilde{n}_{-}\,-\sqrt{\Omega ^{2}\tilde{n}_{-}^{2}+m^{2}%
}$%
\end{tabular}%
\right. .  \label{e16.5}
\end{equation}%
Consequently, one may rearrange such energy levels and cast them so that%
\begin{equation}
E_{\pm }^{\left( \Omega _{+}\right) }=\pm \left\vert \Omega \right\vert \,%
\tilde{n}_{\pm }\pm \sqrt{\Omega ^{2}\tilde{n}_{\pm }^{2}+m^{2}},
\label{e16.6}
\end{equation}%
for positive vorticity, and%
\begin{equation}
E_{\pm }^{\left( \Omega _{-}\right) }=\pm \left\vert \Omega \right\vert \,%
\tilde{n}_{\mp }\pm \sqrt{\Omega ^{2}\tilde{n}_{\mp }^{2}+m^{2}}.
\label{e16.7}
\end{equation}%
for negative vorticity. Notably, we observe that $\tilde{n}_{\pm }\left(
\ell =\pm \ell \right) =\tilde{n}_{\mp }\left( \ell =\mp \ell \right) $
which would in effect introduce the so called vorticity-energy correlations
so that $E_{\pm }^{\left( \Omega _{+}\right) }\left( \ell =\pm \ell \right)
=E_{\pm }^{\left( \Omega _{-}\right) }\left( \ell =\mp \ell \right) $. We
have, therefore, four branches of energy levels so that the upper half
(above $E=0$ line) is represented by $E_{+}$ and the lower half (below $E=0$
line) is represented by $E_{-}$ in the correlations mentioned above. Yet for
massless KG-G\"{u}rses oscillators we obtain $E_{\pm }^{\left( \Omega
_{+}\right) }=\pm 2\left\vert \Omega \right\vert \,\tilde{n}_{\pm }$ and $%
E_{\pm }^{\left( \Omega _{-}\right) }=\pm 2\left\vert \Omega \right\vert \,%
\tilde{n}_{\mp }$.

Moreover, in Figures 1(a) and 1(b) we observe yet a new type of degeneracies
in each branch of the energy levels (i.e., in each quarter of the figures).
That is, states with the irrational quantum number $\tilde{n}%
_{+}=2n_{r}+\left\vert \ell \right\vert +\ell \,+1$ collapse into $\ell =0$
state for $\forall \ell =-\left\vert \ell \right\vert $ and states with $%
\tilde{n}_{-}=2n_{r}+\left\vert \ell \right\vert -\ell \,+1$ collapse into $%
\ell =0$ state for $\forall \ell =+\left\vert \ell \right\vert $. This type
of degeneracies is introduced by the structure of spacetime (G\"{u}rses
spacetime is used here) and therefore should be called, hereinafter,
spacetime associated degeneracies (STADs).

\section{KG-G\"{u}rses plus Mirza-Mohadesi's oscillators}

We now consider KG-G\"{u}rses plus Mirza-Mohadesi's oscillators with $\eta
\neq 0$, and $g\left( r\right) =1$. In this case, our KG-equation (\ref{e11}%
) collapses again into the two-dimensional Schr\"{o}dinger oscillator%
\begin{equation}
R^{\prime \prime }\left( r\right) +\left[ \lambda -\frac{\left( \ell
^{2}-1/4\right) }{r^{2}}-\tilde{\omega}^{2}r^{2}\right] R\left( r\right) =0,
\label{e17}
\end{equation}%
which admits exact textbook solvability so that the eigenvalues and radial
eigenfunctions, respectively, read%
\begin{equation}
\lambda =2\left\vert \tilde{\omega}\right\vert \left( 2n_{r}+\left\vert \ell
\right\vert +1\right) =2\left\vert \Omega E\right\vert \sqrt{1+\frac{\eta
^{2}}{\Omega ^{2}E^{2}}}\left( 2n_{r}+\left\vert \ell \right\vert +1\right) 
\label{e18}
\end{equation}%
and 
\begin{equation}
R\left( r\right) \sim r^{\left\vert \ell \right\vert +1/2}\exp \left( -\frac{%
\left\vert \tilde{\omega}\right\vert r^{2}}{2}\right) L_{n_{n}}^{\left\vert
\ell \right\vert }\left( \left\vert \tilde{\omega}\right\vert r^{2}\right)
\Longleftrightarrow \psi \left( r\right) \sim r^{\left\vert \ell \right\vert
}\exp \left( -\frac{\left\vert \tilde{\omega}\right\vert r^{2}}{2}\right)
L_{n_{n}}^{\left\vert \ell \right\vert }\left( \left\vert \tilde{\omega}%
\right\vert r^{2}\right) .  \label{e19}
\end{equation}%
Then, equation (\ref{e13}) along with (\ref{e18}) imply%
\begin{equation}
E^{2}-2\Omega E\ell -2\left\vert \Omega E\right\vert \sqrt{1+\frac{\eta ^{2}%
}{\Omega ^{2}E^{2}}}\left( 2n_{r}+\left\vert \ell \right\vert +1\right)
-\left( m^{2}+2\eta \right) =0.  \label{e20}
\end{equation}%
It is obvious that for $\eta =0$ in (\ref{e20}) one would exactly obtain the
results for the KG-G\"{u}rses oscillators discussed above. In Figure 2(a),
we notice that the vorticity-energy correlations as well as STADs are now
only partially valid because of the energy shifts introduced by
Mirza-Mohadesi's \cite{Mirza 2004} parameter $\eta $. In Figures 2(b) and
2(c) we can clearly observe such shifts in each quarter of the figures. That
is, quarters 1 and 2 are for $\Omega =\Omega _{+}=+\left\vert \Omega
\right\vert $ (i.e., for $E_{\pm }^{\left( \Omega _{+}\right) }$), and 3 and
4 are for $\Omega =\Omega _{-}=-\left\vert \Omega \right\vert $ (i.e., for $%
E_{\pm }^{\left( \Omega _{-}\right) }$). 

At this point, it should be pointed out that this equation was improperly
treated by Ahmed \cite{Ahmed1 2019}, as he expressed the energies in terms
of $\tilde{\omega}$ where $\tilde{\omega}=\sqrt{\Omega ^{2}E^{2}+\eta ^{2}}$
(see (16) vs (21) with (22) and (16) vs (35) with (36) of \cite{Ahmed1 2019}%
). That is, the energies are given in terms of the energies and his results
starting form his equation (21) to the end of his paper are rendered
misleading, and are incorrect. His results should be redirected to the
results reported in current note, therefore.%
\begin{figure}[!ht!]  
\centering
\includegraphics[width=0.3\textwidth]{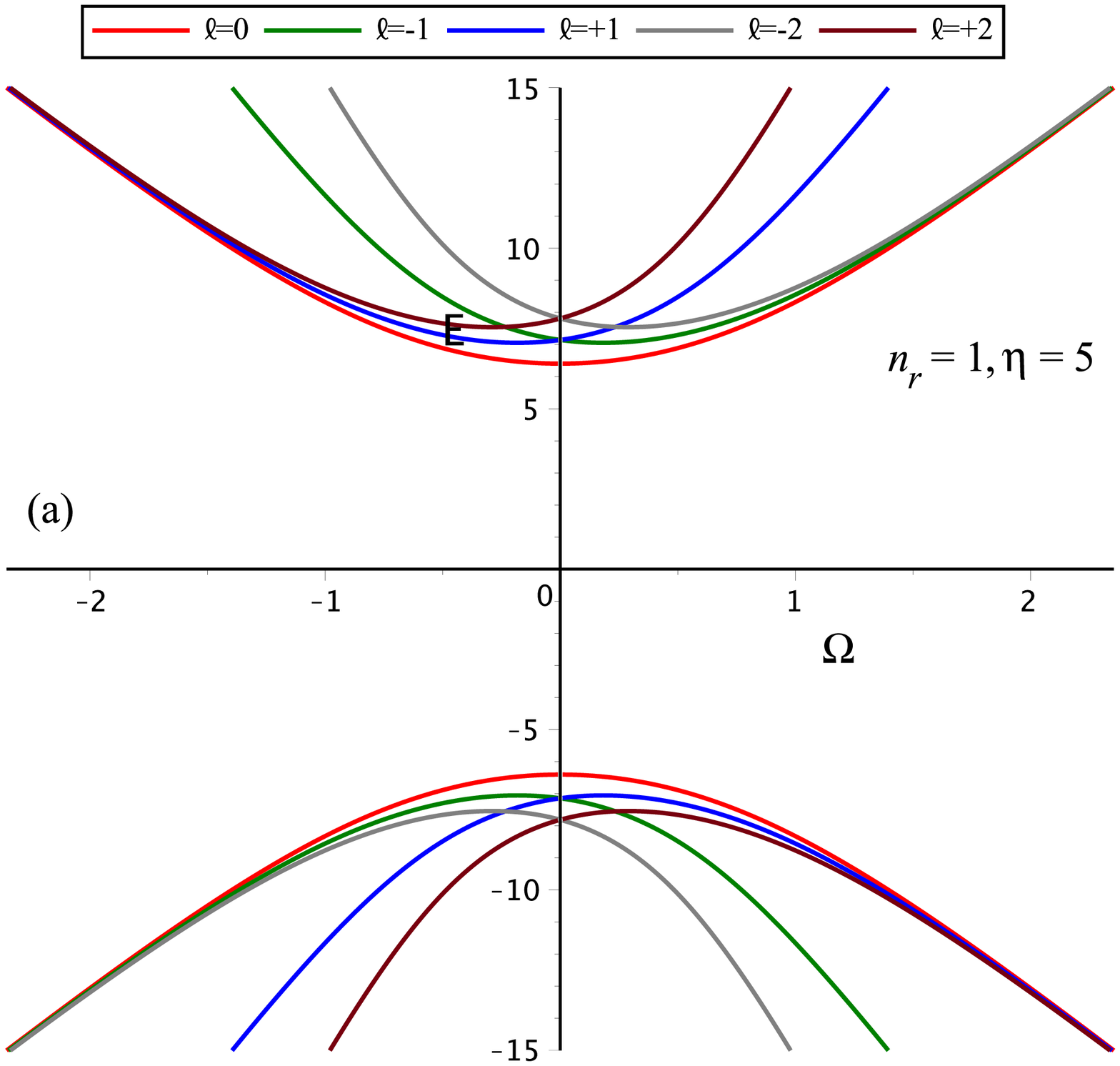}
\includegraphics[width=0.3\textwidth]{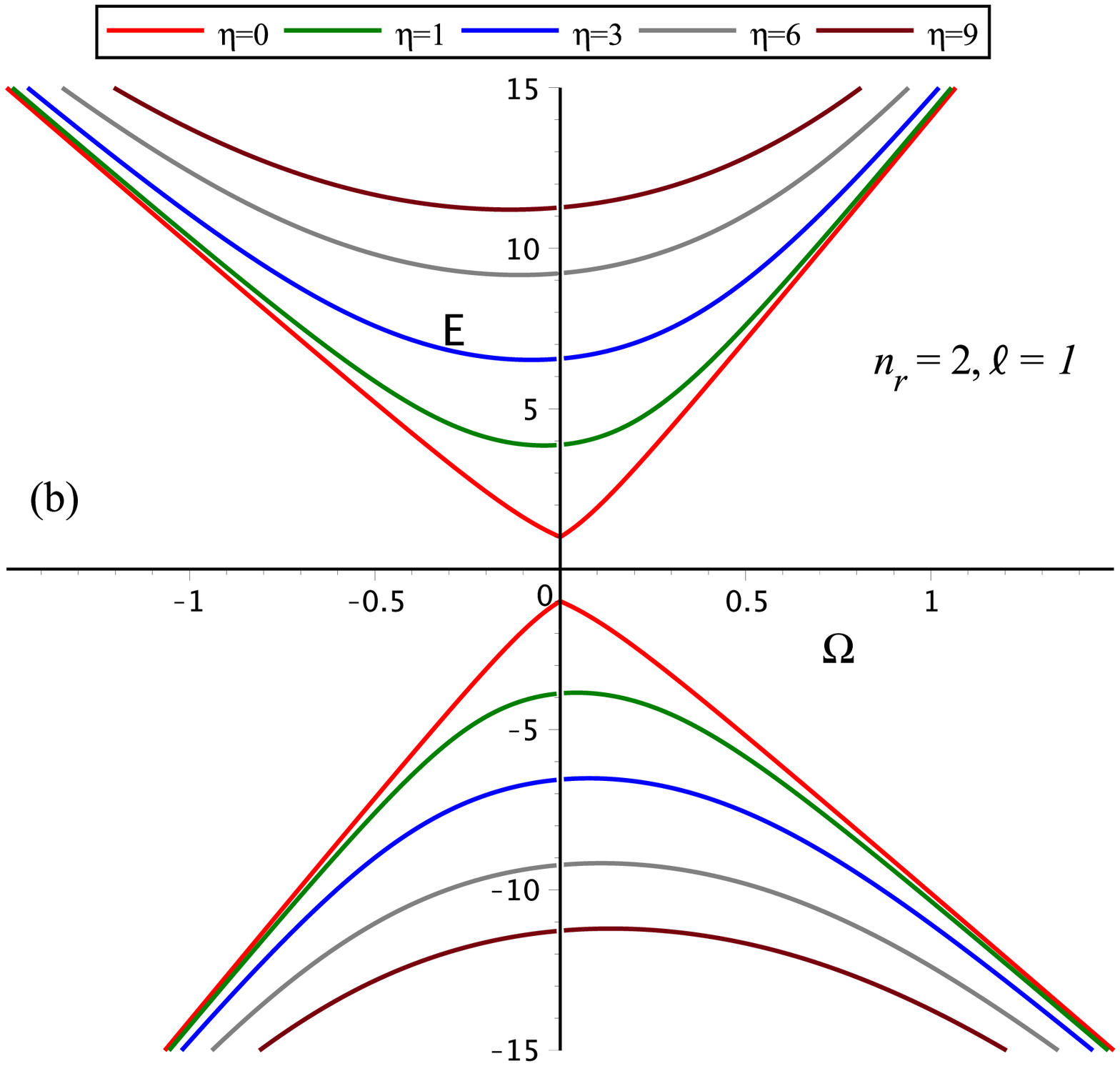} 
\includegraphics[width=0.3\textwidth]{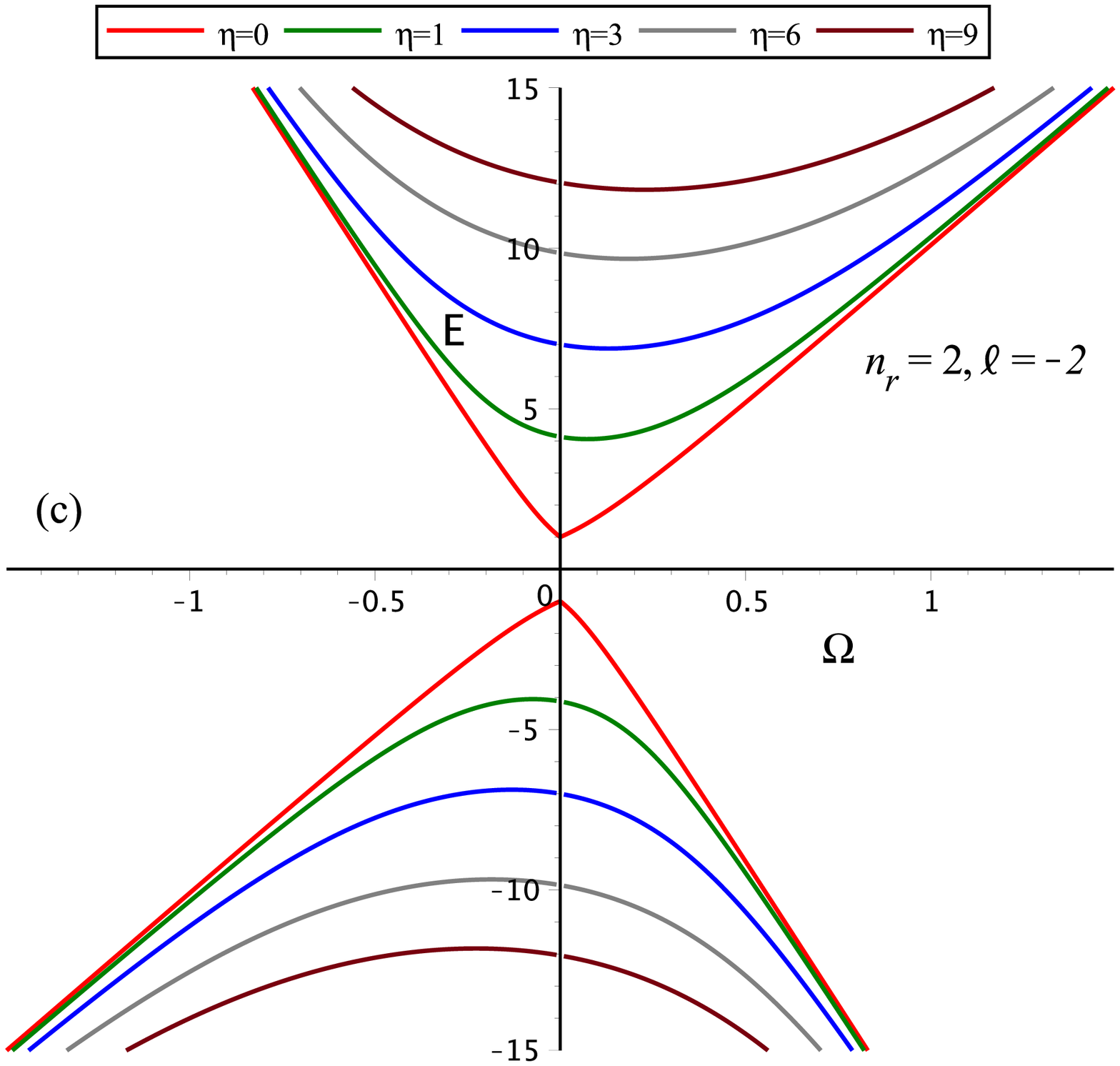}
\caption{\small 
{ The energy levels for KG-G\"{u}rses oscillators of (\ref{e20}) are plotted with $ m=1$ (a) for $\eta=5$, $n_r=1$, $\ell=0, \pm1, \pm2$, (b) for $n_{r}=2$, $\ell=1$, $\eta=0,1,3,6,9$ and (c) for $n_{r}=2$, $\ell=-2$, $\eta=0,1,3,6,9$.}}
\label{fig2}
\end{figure}%

\section{PDM KG-G\"{u}rses oscillators}

In this section we consider PDM settings for KG-G\"{u}rses oscillators,
where $g\left( r\right) =\exp \left( 2\beta r^{2}\right) ;\;\beta \geq 0$.
Under such settings, KG-equation (\ref{e11}) reads%
\begin{equation}
R^{\prime \prime }\left( r\right) +\left[ \lambda -\frac{\left( \ell
^{2}-1/4\right) }{r^{2}}-\tilde{\Omega}^{2}r^{2}\right] R\left( r\right) =0,
\label{e21}
\end{equation}%
with%
\begin{equation}
\lambda =E^{2}-2\,\Omega \,\ell \,E-2\beta -m^{2}\text{ ; \ }\tilde{\Omega}%
^{2}=\Omega ^{2}E^{2}+\beta ^{2}.  \label{e22}
\end{equation}%
In this case, the eigenvalues and radial wavefunctions, respectively, read%
\begin{equation}
\lambda =2\left\vert \tilde{\Omega}\right\vert \left( 2n_{r}+\left\vert \ell
\right\vert +1\right) =2\left\vert \Omega E\right\vert \sqrt{1+\frac{\beta
^{2}}{\Omega ^{2}E^{2}}}\left( 2n_{r}+\left\vert \ell \right\vert +1\right) ,
\label{e23}
\end{equation}%
and 
\begin{equation}
R\left( r\right) \sim r^{\left\vert \ell \right\vert +1/2}\exp \left( -\frac{%
\left\vert \tilde{\Omega}\right\vert r^{2}}{2}\right) L_{n_{n}}^{\left\vert
\ell \right\vert }\left( \left\vert \tilde{\Omega}\right\vert r^{2}\right)
\Longleftrightarrow \psi \left( r\right) \sim r^{\left\vert \ell \right\vert
}\exp \left( -\frac{\left\vert \tilde{\Omega}\right\vert r^{2}}{2}\right)
L_{n_{n}}^{\left\vert \ell \right\vert }\left( \left\vert \tilde{\Omega}%
\right\vert r^{2}\right) .  \label{e23.1}
\end{equation}%
Consequently, the energies are given by%
\begin{equation}
E^{2}-2\,\Omega \,\ell \,E-2\beta -m^{2}\text{ }=2\left\vert \Omega
E\right\vert \sqrt{1+\frac{\beta ^{2}}{\Omega ^{2}E^{2}}}\left(
2n_{r}+\left\vert \ell \right\vert +1\right) .  \label{e24}
\end{equation}%
Obviously, the effect of $\beta $ on the energy levels is the same as that
of \ the Mirza-Mohadesi's oscillators \cite{Mirza 2004} parameter $\eta $.
This would suggest that Mirza-Mohadesi's oscillators \cite{Mirza 2004} may
very well be considered as a special case of PDM KG-oscillators.

\section{KG pseudo-G\"{u}rses oscillators: vorticity-energy correlations and spacetime associated degeneracies}

We now consider a spacetime described by the metric%
\begin{equation}
ds^{2}=-dt^{2}+g\left( r\right) \,dr^{2}-2\Omega Q\left( r\right)
r^{2}dtd\theta +Q\left( r\right) r^{2}\left( 1-\Omega ^{2}Q\left( r\right)
r^{2}\right) d\theta ^{2}.  \label{e25}
\end{equation}%
Next, let us introduce a transformation of the radial part so that%
\begin{equation}
\rho =\sqrt{Q\left( r\right) }r=\int \sqrt{g\left( r\right) }dr\Rightarrow 
\sqrt{g\left( r\right) }=\sqrt{Q\left( r\right) }\left[ 1+\frac{Q^{\prime
}\left( r\right) }{2Q\left( r\right) }r\right] ,  \label{e26}
\end{equation}%
where $%
\mathbb{R}
\ni \left( \rho ,r\right) \in \left[ 0,\infty \right] $, and hence $Q\left(
r\right) \in 
\mathbb{R}
$ is a positive-valued dimensionless scalar multiplier (so is $g\left(
r\right) $). In this case, our spacetime metric (\ref{e25}) now reads%
\begin{equation}
ds^{2}=-dt^{2}+\,d\rho ^{2}-2\Omega \rho ^{2}dtd\theta +\rho ^{2}\left(
1-\Omega ^{2}\rho ^{2}\right) d\theta ^{2}.  \label{e27}
\end{equation}%
This metric looks very much like that of G\"{u}rses (\ref{e1}) and
consequently the KG-equation (\ref{e14}) that describes KG-G\"{u}rses
oscillators is indeed invariant and isospectral with the corresponding KG
pseudo-G\"{u}rses oscillators equation%
\begin{equation}
R^{\prime \prime }\left( \rho \right) +\left[ \lambda -\frac{\left( \ell
^{2}-1/4\right) }{\rho ^{2}}-\Omega ^{2}E^{2}\rho ^{2}\right] R\left( \rho
\right) =0.  \label{e27.11}
\end{equation}
Hence, our KG pseudo-G\"{u}rses oscillators would copy the same\ energies
for the KG-G\"{u}rses oscillators of (\ref{e16.6}) and (\ref{e16.7})
(discussed in section 3) so that%
\begin{equation}
E_{\pm }^{\left( \Omega _{+}\right) }=\pm \left\vert \Omega \right\vert \,%
\tilde{n}_{\pm }\pm \sqrt{\Omega ^{2}\tilde{n}_{\pm }^{2}+m^{2}},
\label{e27.1}
\end{equation}%
for positive vorticity, and%
\begin{equation}
E_{\pm }^{\left( \Omega _{-}\right) }=\pm \left\vert \Omega \right\vert \,%
\tilde{n}_{\mp }\pm \sqrt{\Omega ^{2}\tilde{n}_{\mp }^{2}+m^{2}}.
\label{e27.2}
\end{equation}%
for negative vorticity. However, the radial wavefunctions are now given by 
\begin{equation}
R\left( \rho \right) \sim \rho ^{\left\vert \ell \right\vert +1/2}\exp
\left( -\frac{\left\vert \Omega E\right\vert \rho ^{2}}{2}\right)
L_{n_{n}}^{\left\vert \ell \right\vert }\left( \left\vert \Omega
E\right\vert \rho ^{2}\right) \Longleftrightarrow \psi \left( \rho \right)
\sim \rho ^{\left\vert \ell \right\vert }\exp \left( -\frac{\left\vert
\Omega E\right\vert \rho ^{2}}{2}\right) L_{n_{n}}^{\left\vert \ell
\right\vert }\left( \left\vert \Omega E\right\vert \rho ^{2}\right) .
\label{e27.3}
\end{equation}

The following notes on our spacetime metric (\ref{e25}) are unavoidable.

\begin{description}
\item[(a)] The spacetime metric (\ref{e27}) looks very much like G\"{u}rses
spacetime one of (\ref{e1}) and should be called, hereinafter, pseudo-G\"{u}rses spacetime, therefore.

\item[(b)] If we set $\Omega =-\mu /3$, $a_{_{0}}=1$ in 
\begin{equation}
\phi =a_{_{0}},\,\psi =b_{_{0}}+\frac{b_{_{1}}}{\rho ^{2}}+\frac{3\lambda
_{_{0}}}{4}\rho ^{2},\,q=c_{_{0}}+\frac{e_{_{0}}\mu }{3}\rho
^{2},\,h=e_{_{0}}\rho ,\,\lambda _{_{0}}=\lambda +\frac{\mu ^{2}}{27},
\label{e28}
\end{equation}%
of (\ref{e3}) and use%
\begin{equation}
Q\left( r\right) =e_{_{0}}+\frac{3c_{_{0}}}{\mu r^{2}}\Longleftrightarrow
g\left( r\right) =\frac{\mu e_{_{0}}^{2}r^{2}}{\mu e_{_{0}}r^{2}+3c_{_{0}}},
\label{e29}
\end{equation}%
(where the parametric values are adjusted so that $\left( Q\left( r\right)
,g\left( r\right) \right) \in 
\mathbb{R}
$ are positive-valued functions, i.e., $c_{_{0}}<0$ ) we obtain%
\begin{equation}
q=c_{_{0}}+\frac{e_{_{0}}\mu }{3}r^{2},\;\psi =\frac{1}{e_{_{0}}}+\frac{%
3c_{_{0}}}{\mu e_{_{0}}^{2}r^{2}}.  \label{e30}
\end{equation}%
Which is yet another feasible structure for the G\"{u}rses spacetime of (\ref{e2}) and (\ref{e3}) with%
\begin{equation}
b_{_{0}}=\frac{1}{e_{_{0}}},\,b_{_{1}}=\frac{3c_{_{0}}}{\mu e_{_{0}}^{2}},%
\,\lambda _{_{0}}=0,\text{ }h=e_{_{0}}r.  \label{e31}
\end{equation}

\item[(c)] As long as condition (\ref{e26}) is satisfied, all KG-pseudo-G\"{u}rses oscillators (including the one in (b) above) in the spacetime model of (\ref{e27}) admit isospectrality and invariance with the KG-G\"{u}rses oscillators (\ref{e14}) and inherit the same vorticity-energy correlations so that $E_{\pm }^{\left( \Omega _{+}\right) }\left( \ell =\pm
\ell \right) =E_{\pm }^{\left( \Omega _{-}\right) }\left( \ell =\mp \ell
\right) $ as well as they inherit the spacetime associated degeneracies,
discussed in section 3.
\end{description}

\section{Concluding remarks}

In the current proposal, we revisited KG-oscillators in the
(1+2)-dimensional G\"{u}rses spacetime of (\ref{e1}) so that PDM settings
and Mirza-Mohadesi's KG-oscillators \cite{Mirza 2004} are included. We have
observed that  KG-G\"{u}rses oscillators are introduced as a byproduct of
the very nature of the G\"{u}rses spacetime structure. This has, in turn,
motivated us to first elaborate and discuss the effects of G\"{u}rses
spacetime on the energy levels of the KG-G\"{u}rses oscillators. We have
found that such KG-G\"{u}rses oscillators admit vorticity-energy
correlations as well as spacetime associated degeneracies (STADs)
(documented in Figures 1(a) and 1(b)).. However, for KG-G\"{u}rses plus
Mirza-Mohadesi's oscillators we have observed that the vorticity-energy
correlations as well as STADs are only partially valid because of the energy
shifts introduced by Mirza-Mohadesi's \cite{Mirza 2004} parameter $\eta $
(documented in Figures 2(a), 2(b), and 2(c)). Nevertheless, this model was
studied by Ahmed  \cite{Ahmed1 2019} who has reported improper treatment and
incorrect results. Consequently, his reported results (starting from his
equation (21) to the end of his paper) should be redirected to the ones
reported in the current study. Moreover, we have shown that PDM setting may
very well have the same effect on the spectrum as that reported for KG-G\"{u}rses plus Mirza-Mohadesi's oscillators. Yet, a new set of the so called KG pseudo-G\"{u}rses oscillators is introduced and is shown to be invariant and
isospectral with KG-G\"{u}rses oscillators. Therefore, such KG pseudo-G\"{u}rses-oscillators would inherit the vorticity-energy correlations as well as STADs of the KG-G\"{u}rses oscillators.

\textbf{Data Availability Statement} Authors can confirm that all relevant data are included in the article and/or its supplementary information files. The author confirms that there are no online supplementary files (for web only publication) in this article.

\end{document}